\documentclass[aps,pre,twocolumn,showpacs,superscriptaddress]{revtex4}

\usepackage{graphicx} 

\usepackage[dvips, colorlinks=true, citecolor=blue, urlcolor=blue, linkcolor=blue, ]{hyperref}

\usepackage{amssymb,amsmath}

\usepackage{bbm}

\usepackage{float}

\begin{document}


\title{Coagulation and fragmentation dynamics of inertial particles}

\author{Jens C. Zahnow}

\affiliation{Theoretical Physics/Complex Systems, ICBM, University of Oldenburg, 26129 Oldenburg, Germany}

\author{Rafael D. Vilela}

\affiliation{Max Planck Institute for the Physics of Complex Systems, 01187 Dresden, Germany}
\affiliation{CMCC, Universidade Federal do ABC, 09210-170 Santo Andr\'e, SP, Brazil}

\author{Ulrike Feudel}

\affiliation{Theoretical Physics/Complex Systems, ICBM, University of Oldenburg, 26129 Oldenburg, Germany}

\author{Tam\'as T\'el}

\affiliation{Institute for Theoretical Physics, E\"otv\"os University, H-1518 Budapest, Hungary}



\begin{abstract}

Inertial particles suspended in many natural and industrial flows
  undergo coagulation upon collisions and fragmentation if their size
  becomes too large or if they experience large shear.  Here we study
  this coagulation-fragmentation process in time-periodic
  incompressible flows.  We find that this process approaches an
  asymptotic, dynamical steady state where the average number of
  particles of each size is roughly constant. We compare the
  steady-state size distributions corresponding to two
  fragmentation mechanisms and for different flows and find that the steady state is mostly independent of the coagulation process. While collision rates determine the transient behavior, fragmentation determines the steady state. For example, for fragmentation due to shear, flows that have very different local particle concentrations can result in similar particle size distributions if the temporal or spatial variation of shear forces is similar. 

\end{abstract}

\maketitle

\section{Introduction}

\label{sec:introduction}
The dynamics of inertial particles in fluid flows plays an important role in
many natural and industrial contexts and has received an increasing interest in recent years. Questions where inertial particles play an important role are ubiquitous in biology, chemistry, oceanography, astro- and geophysics. Recent works from dynamical systems
\cite{Nishikawa2001,Benczik2002,Lopez2002,Cartwright2002,Do2004,Benczik2006,Vilela2007,
Sapsis2008} to atmospheric science \cite{Shaw2003,Falkovich2007,Jaczewski2005} and
turbulence \cite{Wilkinson2005,Bec2005,Calzavarini} have added greatly to the understanding of these phenomena. Almost all these works have
been devoted to the dynamics of non-interacting inertial particles. A major
reason for this is that this problem is already very rich,
displaying features yet to be understood in their full complexity, as
inhomogeneous spatial distributions \cite{Maxey1986,Maxey1987a} and multivalued velocity
fields \cite{falko_nat02,Wilkinson2005,Wilkinson2007}. Interestingly, these very same
features yield an increased rate of collisions \cite{Bec2005}, the consequences of
which are in most cases not explicitly taken into account. Typically one
assumes a {\em dilute regime} and fully neglects the collisions. In some other
works, one keeps track of the collisions numerically without actually
addressing the outcome of such events ({\em ghost collisions})
\cite{Wang2000,Zhou2001,Bec2005}. To our knowledge, only very recent works have
addressed effects of collisions on the dynamics of inertial particles
\cite{Wilkinson2008,Zahnow2008_2,Medrano2008}. In Ref.~\cite{Zahnow2008_2}, we have
reported our first results on the dynamics of inertial particles
coagulating \footnote{In Ref.~\cite{Zahnow2008_2} coagulation was referred to as aggregation. However, in the context of liquid particles coagulation is the more widely used term.} upon collisions and fragmenting under certain
conditions. In Ref.~\cite{Medrano2008}, the authors considered elastic
collisions in a monodisperse system and pointed out the existence of
bursts in the spread of the particles out of the attractors of
the purely advective dynamics. In Ref.~\cite{Wilkinson2008} the authors treated coagulation and shear-fragmentation of dust particles in an astrophysical context. There small dust particles can grow into larger fractal clusters due to turbulent collisions.

In this paper we extend the work of Ref. \cite{Zahnow2008_2} to different
flows and to a broadened parameter set. Our motivation lies primarily on
natural phenomena such as the collisional growth of cloud droplets
\cite{Pruppacher1997}, sediments in lakes and rivers, and marine snow in the
ocean \cite{Alldredge1990}. Here we focus on the description of spherical droplets, i.e. we do not take into account any fractal structures that often appear in sediments or marine snow. 

Our main result is that coagulation and fragmentation dominate the behavior of different time spans of the process and fragmentation rather than coagulation is the dominating process for the steady state size distribution. For different flows, the collision rates between inertial particles can be very different, leading to great changes in the coagulation of particles. While this might be an important effect for transient processes, such as the initiation of rain in clouds, it will turn out that for the steady state fragmentation plays a much greater role. In fact, the steady state size distributions are mainly determined by the fragmentation process. The specific flow structure is only relevant for the steady state size distribution when it directly affects the fragmentation process. This can for example be the case for fragmentation due to shear when the spatial and temporal variation of the shear is very different for two flows.

We consider fragmentation to be of two possible
origins. First, particles break up if their size exceeds a certain maximum
allowed size.  This is motivated by the hydrodynamical instability of large
water drops (e.g. cloud drops) settling due to gravity
\cite{Viller2008}. Second, particles fragment if the shear forces due to the
fluid flow are sufficiently large. This mechanism has been reported to be the
dominant one in the case of marine aggregates \cite{Thomas1999}.

At a first glance, one might be tempted to pursue a field-theoretical
approach, in the framework of which one treats the problem
of particle motion as a multiphase flow and then applies the Smoluchowski
equation \cite{Smoluchowski1917} to model coagulation and fragmentation for
the particle distribution. However, the inertial particle dynamics is dissipative and contracts to an attractor in a $2d$-dimensional phase space, where $d$ is the spatial dimension of the flow. This attractor can be folded in phase space, meaning that the particle velocity may take on several
values even at the same location. Due to the presence of such 'caustics' \cite{falko_nat02,Bec2003,Wilkinson2005}, a field-theoretical approach cannot be well founded. Therefore a study based on an individual tracking of the particles, as the one presented here, becomes necessary.

Here we consider the fluid flow to be spatially smooth and to have a single macroscopic time scale.
We are motivated by flows having coherent (e.g. convective) structures
on length scales much larger than the ones at which turbulence plays a
major role.  The effect of turbulence can then be taken into account
as a stochastic perturbation described by an eddy diffusivity
\cite{Kundu} at small scales. For simplicity, we neglect this small
scale noise in the present work and focus only on the large scale
motion of the fluid.

We study the dynamics of the system formed by the transported inertial particles
undergoing coagulation and fragmentation in  three different fluid flows,
as described in Section II.  We find that the system tends to approach
a steady state where several size classes coexist (Section III).  The
average number of particles in each size class is roughly constant with a
mild periodic time dependence --- with a period identical to the one of the
advecting fluid flow. The distribution of particles as well as the mean average size in the steady state
depends on the type of fragmentation mechanism taking place. First, when
fragmentation occurs solely due to particles exceeding a maximum allowed size,
the distribution is in general quite broad. Second, for fragmentation
occurring also under sufficiently large shear, the distributions typically
decay exponentially beyond a certain size class.  The distributions
depend on the fluid flow for both types of fragmentation. However, for shear fragmentation the differences are very small as long as the variation of the fluid shear over time is qualitatively similar.  

In the case of shear fragmentation, we derive a scaling relation for the average size class
in the steady state as a function of the coagulate strength parameter
$\gamma$. Finally, we show that our results are robust with respect to the total mass of particles, the number of allowed size classes and the initial particle size distribution. Also, in the case of shear fragmentation the size distribution in the steady state has a scaled functional form which does not depend on the coagulate strength $\gamma$.

\section{Coagulation and Fragmentation Model}

\label{sec:aggfrag_model}

\subsection{Dynamics of Inertial Particles}

\label{subsec:advection}

First, we present the equations of motion for the motion of finite size particles that will be used here. For simplicity we consider heavy spherical aerosols, i.e. particles much denser than the ambient fluid and assume that the difference between their velocity $\dot{\bf x}$ and the fluid velocity  ${\bf u}={\bf u}({\bf x}(t),t)$ at the same position is sufficiently small so that the drag force is proportional to this difference (Stokes drag). The dimensionless form of the governing equation for the path ${\bf{x}}(t)=(x_1(t),x_2(t))$ of the center of mass for such heavy aerosols subjected to drag and gravity reads in this case as \cite{Maxey1983,Auton1988, Michaelides1997}:
\begin{equation}
\ddot{\bf x}=\frac{1}{\tau}\left({\bf u}({\bf x}(t),t)-\dot{\bf x}-W{\bf n}\right), 
\label{eq:maxey}
\end{equation}
where ${\bf n}$ is a unit vector pointing upwards in the vertical direction. Throughout this paper we consider the vertical direction along the axis $x_2$. Under the assumption that the density ratio ${\rho_f}/{\rho_p}\ll 1$, the particle response time $\tau$ can be written in terms of the density  $\rho_p$ of the particle, the radius $r$ of the aerosols, the fluids dynamic viscosity $\eta$, and the characteristic length $L$ and characteristic velocity $U$ of the flow as $\tau=(2r^2\rho_pU)/(9\eta L)$. 
We note that the response time $\tau$ is nothing but the
Stokes number which can be written in our case as $\tau=\tau_p/T$
where $\tau_p$ is the particle's dimensional Stokesian relaxation time and $T$ is  the
characteristic time of the flow.
The dimensionless settling velocity in a medium at rest is given by $W=(2r^2 \rho_p g)/(9 \eta U)$. Note that $W/\tau$ is independent of the particle radius $r$.

Every particle produces perturbations in the flow that decay at least inversely proportional to the distance from the particle \cite{Happel1983,JTWG1997}. Here we assume a dilute regime, where the local concentration of particles is low enough, so that particle-particle interaction can be neglected \cite{concentration} unless particles come into direct contact.

The assumption that the particle radii $a$ are small also means that the feedback from the particle motion on the flow will be small as well \cite{Michaelides1997} and is therefore neglected in the following. 

\subsection{Coagulation}
\label{subsec:coagulation}

Second, we present a model for the coagulation of finite size particles. 

The smallest particles considered will be called
\textit{primary} particles. These primary particles can combine to form larger
particles, called \textit{coagulates}. Coagulation takes place upon
collision. All particles are assumed to consist of an integer number of these
primary particles, i.e. the primary particles can never be broken up. The
number $\alpha$ of primary particles in a coagulate is called the \textit{size
class index}. We consider $n$ different size classes, i.e. coagulates can
consist of a maximum of $n$ primary particles. A coagulate of size class
$\alpha$ has a radius $r_{\alpha}=\alpha^{1/3}r_{1}$, where $r_1$ is the
radius of the primary particles. The response time is $\tau_{\alpha}=
(r_{\alpha}/r_1)^2 \tau_{1}=\alpha^{2/3} \tau_1$ and the settling velocity in
still fluid is $W_{\alpha}= {\alpha}^{2/3} W_{1}$. Here $\tau_1$ and $W_1$ are
the response time and the settling velocity for the primary particles,
respectively. The largest coagulates therefore have a radius
$r_{n}=n^{1/3}r_1$. We note that particles of different sizes have different
parameters $\tau_\alpha$ and $W_\alpha$ and therefore follow the flow with
different parameters in the equation of motion (\ref{eq:maxey}).

We define a collision of two particles if the centers of the particles, say of
radius $r_i$ and $r_j$, come closer than a distance $d=r_i+r_j$. In that case
the particles coagulate and form a larger particle. Mass conservation requires
the radius of the new particle to be $r^3_{new}=r^3_i+r^3_j$. For the size
class index this implies a linear rule, $\alpha_{new}=\alpha_i+\alpha_j$, which
determines the new response time and settling velocity via
$\tau_{\alpha_{new}}=\alpha_{new}^{2/3} \tau_1$ and
$W_{\alpha_{new}}=\alpha_{new}^{2/3} W_1$, respectively.

The velocity of the new particle follows from momentum conservation. The
position of the new coagulate is the center of gravity of the two old particles.

\subsection{Fragmentation}
\label{subsec:fragmentation}
Third, we present a model for the fragmentation of particles. Primary particles cannot be broken up. In the following, we will compare two different fragmentation rules.

(i) \textit{Size-limiting fragmentation}: If a particle becomes larger than the maximum radius $a_{n}$, it is broken up into two smaller fragments (binary fragmentation) whose radii are chosen randomly, from a uniform distribution between $a_1$ and half the original radius. If any fragment is larger than $a_{n}$ this process is repeated, until no fragment exceeds $a_{n}$.

(ii) \textit{Shear fragmentation} takes place when the hydrodynamical force $F_{hyd}$ acting on the particle exceeds the forces $F_{coag}$ holding the coagulate together by a certain factor. The criterion for breakup can therefore be expressed as 
\begin{equation} \label{eq:splitting_cond}
F_{hyd}/F_{coag}>\tilde{\gamma}~
\end{equation}
where $\tilde{\gamma}$ is a constant. 

The hydrodynamical force in this case is proportional to the local velocity
gradients in the flow. It is expected that larger particles are more likely to break-up, therefore the critical force required for fragmentation should decrease with the coagulate size. For liquid spherical particles (drops) in the size
range where viscous forces dominate, Taylor \cite{Taylor1934} and later
Delichatsios \cite{Delichatsios1975} derived an expression for the critical
velocity difference $\Delta u$ across the drop required for breakup. Under the condition that the characteristic time of drop deformations is small compared to the time where this velocity gradient
occurs, we rewrite this condition with a single parameter as
 \begin{equation}\label{eq:splitting_drop}
\frac{ \Delta u}{a_{\alpha}} = \gamma \left(\frac{r_1}{r_{\alpha}}\right)=\gamma \alpha^{-1/3}
 \end{equation}
 where $\gamma$ is a constant, the coagulate strength parameter (the same quantity is called stickiness in \cite{Zahnow2008_2}). The radius has been normalized with the radius of a primary particle. If the maximum velocity difference across the radius of the drop exceeds the threshold value given by Eq. \eqref{eq:splitting_drop}, the particle is broken up into two smaller fragments (binary fragmentation) in the same way as for size-limiting fragmentation. 

At the instant of both coagulation and fragmentation there is a sudden change in the dynamics: the number of particles changes in $2$ or $3$ among the $n$ available dynamical systems defined by the size classes. 

\subsection{Fluid flows}
\label{subsec:fluid_flows}
For convenience, we treat the case where the fluid flow depends only on two coordinates, i.e. we study a three-dimensional flow where the velocity in the third direction is negligible compared to the other two velocities. This can then be represented by a two-dimensional flow. Therefore, the phase space of the particles dynamics is four-dimensional. We choose three simple paradigmatic flow situations with different characteristics to indicate the generality of our results. 

All flow domains are spatially periodic, with a characteristic length $L$. More specifically, the flows are (a) a convection cell flow with moving vortex centers (in the following referred to as the \textit{moving convection flow}), (b) a convection cell flow with fixed vortices (referred to as the \textit{fixed convection flow}), (c) a sinusoidal shear flow. 

The two convection cell flows (a) and (b) consist of a regular pattern of vortices, or roll cells. Flow (b) was first introduced by Chandrasekhar \cite{Chandrasekhar} as a solution to the Rayleigh-B\'enard problem and since then it has been used in the context of different theoretical studies \cite{Maxey1987,Nishikawa2001,Nishikawa2002,Zahnow2008}. The moving convection flow (a) is a slightly modified version, with moving vortex centers, to yield a more realistic chaotic regime for the particle motion. Convection flows are chosen because they contain vortices (convection cells) and uprising/sinking regions, which are characteristic features of realistic flows often found in nature.
The flows are defined by the velocity field as\\
(a) moving convection flow
\begin{equation}\label{eq:convection_flow}
{\bf u}(x_1,x_2,t)= [1+k_1\sin(\omega_1 t)]\left(\begin{array}{c}
\sin(2\pi \hat{x_1})\cos(2\pi \hat{x_2}) \\
-\cos(2\pi \hat{x_1})\sin(2\pi \hat{x_2})
 \end{array}\right)~,
 \end{equation}
where $\hat{x_1}=x_1+k_2\sin(\omega_2 t)$ and $\hat{x_2}=x_2+k_2\cos(\omega_2 t)$. The parameters $k_1=2.72$ and $\omega_1=\pi$ are the amplitude and the frequency of the periodic forcing of the flow, respectively. $k_2=1/(2\pi)$ and $\omega_2=\pi/4$ determine the amplitude and the frequency of the periodic motion of the centers of the vortices in the flow. The period of the flow is $T=2$ and the characteristic length and characteristic velocity are $L=1$ and $U=1$.\\
(b) fixed convection flow with the same equation for the flow as in (a), but with $k_2=0$.

The sinusoidal shear flow (c) consists of alternating horizontal and vertical velocity components, where each velocity component consists of two plateaus in time. It was introduced in Refs. \cite{Liu1994,Pierre1994} and has been used many times in chaotic advection studies. Here we consider a time-continuous version (see \cite{Vilela2007_2}) defined by:\\
(c) sinusoidal shear flow
\begin{equation}\label{eq:shear_flow}
{\bf u}(x_1,x_2,t)= 0.5\left(\begin{array}{c}
(1+\tanh(\beta \sin(2\pi t)))\sin(2\pi x_2) \\
(1-\tanh(\beta \sin(2\pi t)))\sin(2\pi x_1)
 \end{array}\right)~,
 \end{equation}
where the parameter $\beta$ describes how rapidly the transition between two values, a zero and a nonzero velocity, takes place
for each velocity component. The typically used value $\beta=20/\pi$
corresponds to a very rapid transition. 

The period of the flow is $T=1$ and the characteristic length and characteristic velocity are $L=1$ and $U=1$.

\begin{figure}[H]
		\centering
		\includegraphics[width=0.46\textwidth]{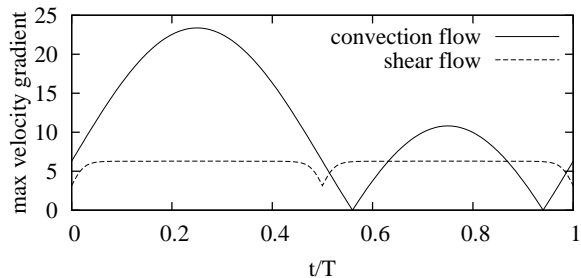}
 		\caption{\label{fig:flow_gradient} Maximum of the velocity gradient grad ${\bf u}(x_1,x_2,t)$ vs. time for the two different flows described by Eqs. \eqref{eq:convection_flow} and \eqref{eq:shear_flow}. }	
\end{figure}

The fluid flows are laminar and time-periodic, but the dynamics of the inertial particles moving in these flows can be chaotic. 

To emphasize the difference between the flows, Fig. \ref{fig:flow_gradient} shows the maximum of the velocity gradient vs.  time for the convection flows (there is no difference between the moving and the fixed convection flow) and the sinusoidal shear flow. The difference in magnitude and also in the temporal evolution between these two flows is clearly visible, indicating a possibly very different behavior with respect to shear fragmentation.

\subsection{Numerical implementation}
\label{subsec:implementation}

After presenting the model, we describe some details about the
implementation. In the bulk of the paper we consider $n=30$ size classes. The
primary particles considered here have dimensionless radius
$r_{1}={5}/{30^{1/3}}\times10^{-5}$, response time $\tau_1=1/55$ and settling velocity $W_{1}=3.2\tau_1$. As initial condition we take $N(t=0)=10^5$ particles of the smallest size and no larger
particles. Furthermore, particles are uniformly distributed over the
$1\times1$ unit cell of the configuration space. The initial particle velocity matches
that of the fluid at their position in all cases.  

The simulation is based on the following ingredients: 

a) All particles move in the flow over some time step $dt$ according to Eq. \eqref{eq:maxey}. This integration time step $dt$ needs to be chosen small enough to allow for the detection of every collision. After each time step $dt$ there is an interaction between particles in the form of coagulation if they are too close to each other. Our experience shows
that a choice $dt=T/20$ is sufficiently small for the conditions considered here.

Because of the spatial periodicity of the flow, the particle dynamics is
folded back onto the $1\times 1$ unit cell, using periodic boundary conditions
(see e.g. \cite{Nishikawa2002,Zahnow2008}).

b) Particles coagulate if their distance is smaller than the sum of their radii.
Computationally, the coagulation process is the most costly component of the simulation. Here a link-cell algorithm \cite{Hockney1981} is used to compute the distance between particles, which scales as $O(N)$ and is thus much faster than simply summing over all particles.

c) Coagulates can fragment either due to size-limiting fragmentation or due to shear fragmentation.
\begin{enumerate}
\item Size-limiting fragmentation: If the coagulate size $\alpha$ exceeds the predefined maximum size, which is in the following fixed at $n=30$ unless mentioned otherwise, the coagulate is broken up.
\item Shear fragmentation: If the shear at the position of the coagulate exceeds a critical value, determined by Eq. \eqref{eq:splitting_drop} the coagulate breaks up. Due to the symmetry of the flows chosen here, the maximum velocity difference is always in the direction of one of the coordinate axes, therefore only these values have to be calculated. Shear fragmentation is always applied together with size-limiting fragmentation to keep the maximum number of occurring size classes fixed at $n$.
\end{enumerate}
Whatever rule is applied, the result is the reversed process of coagulation: two new particles are formed from an old one with the size class indices: $\alpha_{i,new}+\alpha_{j,new}=\alpha_{old}$. 
 As indicated earlier, $\alpha_{i,new}$ can take on any value between 
$\alpha_1$ and $\alpha_{n/2}$ with equal probability. The centers of the new particles are placed along a line segment in a random direction so that the distance $d$ between the particle centers equals the sum of their radii, i.e. $d=r_i+r_j$, and the center of mass remains unchanged. Momentum is conserved. For simplicity we assume that the new particles have the same velocity as the old one.

\section{Simulation results}

\label{sec:sim_results}
In this section we show simulation results using the model described above and compare the influence of the different flows, and the effect of size-limiting fragmentation and shear fragmentation.

\begin{figure}[htb]
		\centering
		`\includegraphics[width=0.5\textwidth]{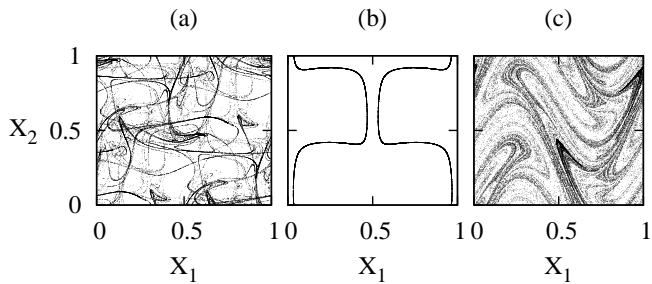}
 		\caption{\label{fig:attraktor1} Stroboscopic sections of the attractors of Eq. \eqref{eq:maxey} projected onto the configuration space for particles for size class $\alpha=16$ for: (a) moving convection flow, (b) fixed convection flow and (c) sinusoidal shear flow. The difference in the degree of clustering in the three flows is clearly visible.}

\end{figure}

Before presenting any results for the complete model, it is worth showing the attractors for the non-interacting problem in the different flows. Fig. \ref{fig:attraktor1} shows the stroboscopic section (taken with the period $T$ of the flow) of the attractors for the flows (a)-(c) for one specific size class projected onto the plane of the coordinates. The figure illustrates the difference in the geometric properties of the particles dynamics in the different flows.

 For the moving convection flow and the sinusoidal shear flow the degree of
 clustering of the particles in the attractors, quantified by their fractal
 dimension, decreases monotonically with the size class. The parameter region
 is chosen in such a way that the attractors are either area filling or 
fractal with dimension smaller than $2$, which we consider to be closer to a realistic
 situation than for example fixed point attractors.  For the fixed convection flow, the particles tend to cluster on a quasiperiodic attractor. This leads to a much larger collision rate than in the other two cases. 

\begin{figure}[htb]
		\centering
		\includegraphics[width=0.46\textwidth]{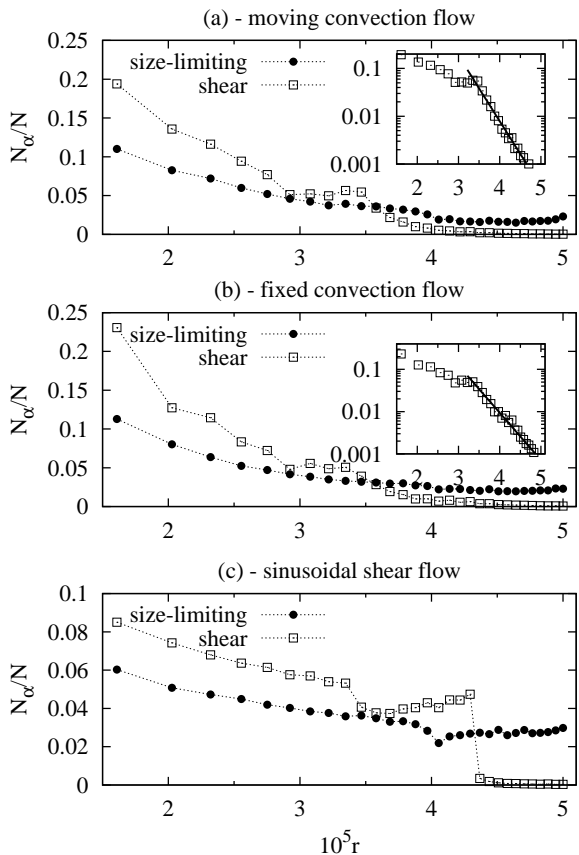}
 		\caption{\label{fig:hist_1}  Histogram of the particle size distribution in steady state for size-limiting fragmentation and shear fragmentation with (a) moving convection cell flow, (with $\gamma=50$ for shear fragmentation), (b) fixed convection cell flow, (with $\gamma=45$ for shear fragmentation), (c) sinusoidal shear flow (with $\gamma=17$ for shear fragmentation). 
The insets show the exponential tail of the size distributions for shear fragmentation in the case of the moving convection cell flow and the fixed convection cell flow. 
}	
\end{figure}

In all flows we find convergence to an asymptotic steady state. Initially, coagulation leads to a fast increase in the average particle size class, independent of the fragmentation rules. Then fragmentation sets in and a balance between coagulation and fragmentation is reached, with an asymptotic average coagulate size
\begin{equation} 
\alpha_\infty=\lim\limits_{t\rightarrow\infty}\frac{1}{T}\int\limits_{t}^{t+T}ds\left<\alpha(s)\right>
\end{equation}
that depends on the fragmentation rule and the different flows. The average $\left<\alpha(s)\right>$ is taken over the coagulate sizes at time $s$
and Eq. (6) corresponds to time-averaging $\left<\alpha(s)\right>$ over
one time period of the flow to remove the periodicity. For the transient behavior of the coagulation-fragmentation process the geometric properties, in particular the degree of clustering of the particles, are very important since they affect the time scales of the transients. This is not the case for the steady state. This becomes very clear when looking at the size distributions of coagulates in steady state for the different flows and fragmentation mechanisms (Fig. \ref{fig:hist_1}). One might expect that the steady state for flow (b) is very different from the other two cases, due to the large difference in particle clustering. Our results show however, that the steady state of the particle dynamics for flows (a) and (b) are almost identical, while flow (c) produces different results. For example for both flows (a) and (b) the size distribution
has a long tail towards larger size classes that decays exponentially for shear fragmentation. Similar exponential tails for the particle size distribution have been found in observations of aggregates in the ocean (see e.g. \cite{Lunau2006}). 

By contrast, for the sinusoidal shear flow the size distribution has two peaks and then drops off sharply towards zero beyond the second peak. In this case the size distribution for shear fragmentation is almost identical to that of size-limiting fragmentation, but for a lower value of $\alpha_{max}$. This is due to the fact that in the sinusoidal shear flow for the chosen parameters the shear a particle experiences is almost constant over time and space, except for a small 'dip' every half period. This very narrow distribution of the shear is very similar to having a single maximum stable size, as is the case for size-limiting fragmentation. In this case a value of $\gamma=17$ for shear fragmentation corresponds to a value of $\alpha_{max}=20$.

We have also checked the size distributions in subregions of the flows. We found, that as long as the number of particles in the subregions allowed sufficiently good statistics, the normalized size distributions coincided with the global distributions. This means that there is no significant spatial dependence of the size distribution. 

While the specific shape of the size distributions found may not be very general, both because of the limitation to only thirty size classes and the very simplified flows, our approach illustrates clearly that the geometric properties of the particle motion related to preferential concentration are not the most relevant ones for the steady state distributions. Instead, the most important effect for the steady state of the particles seems to be the fragmentation process.

\begin{figure}[htb]
		\centering
		\includegraphics[width=0.46\textwidth]{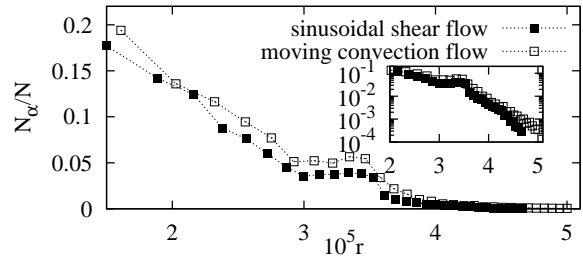}
 		\caption{\label{fig:hist_2} Histogram of the particle size distribution in steady state for the moving convection cell flow ($\gamma=50$) and the smoothed out sinusoidal shear flow ($\gamma=8.5$, $\beta=\pi/20$) for shear fragmentation.}	
\end{figure}

 Size-limiting fragmentation is the same in all flows, as it does not depend on properties of the flow. In this case the differences in the size distributions are small for the different flows. This indicates that the size distributions are mainly determined by the fragmentation process, because the flow specific differences, e.g. differences in coagulation rates, do not affect the shape of the size distribution.  However, when fragmentation depends on shear, the different flows produce very different size distributions. This difference is mainly due to the different properties of the shear forces in the flow (Fig. \ref{fig:flow_gradient}) which lead to differences in fragmentation. To see that it does not depend e.g. on the detailed
characteristics of the particle motion or on the different collision rates, we can adjust the flow
parameters. Decreasing the value of the parameter $\beta$ for the sinusoidal shear flow to
a much smaller one, e.g. $\beta=\pi/20$ greatly increases the variation in the shear forces over space and time. We
obtain two sinusoidal peaks per period for the shear forces, similar to what
happens in the convection flow (cf. Fig. \ref{fig:flow_gradient}), except that
for the sinusoidal shear flow both peaks have the same height.
 For an
appropriate choice of $\gamma$, so that the average size classes match, it can
be seen that the shape of the particle size distributions for both flows have
become almost identical (Fig. \ref{fig:hist_2}) and show the characteristic exponential tail. Again, this indicates that the shape of the size distribution is mostly affected by fragmentation and any flow specific differences in the distribution result mainly from differences in the shear distribution in the flow, which in turn change the fragmentation rates. 

\begin{figure}[htb]
		\centering
		\includegraphics[width=0.46\textwidth]{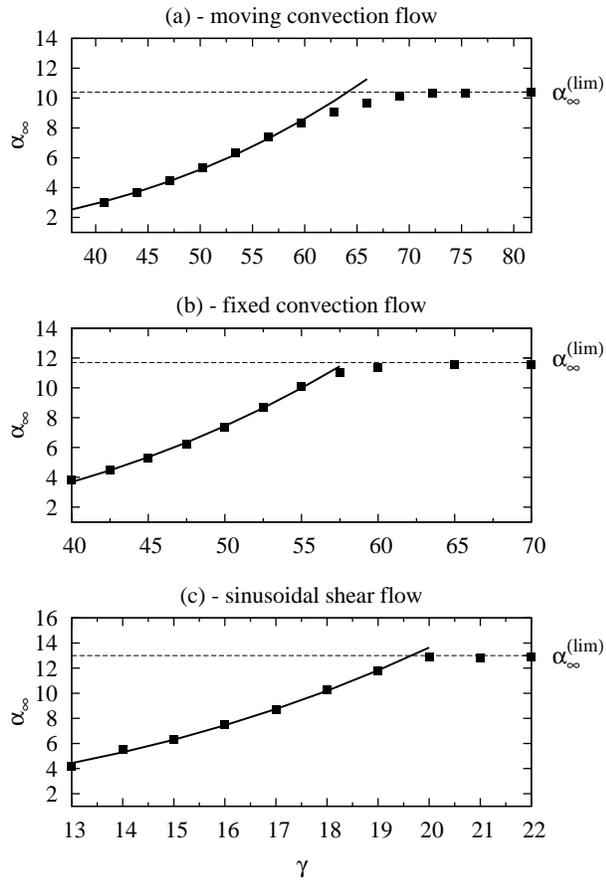}
 		\caption{\label{fig:average_2} Asymptotic average size class  index $\alpha_{\infty}$ as a function of the coagulate strength parameter $\gamma$. Squares: numerical results, solid line: fit by $\alpha_{\infty}=c_1+c_2\gamma^3$ (see Eq. \eqref{eq:alpha_crit}). (a) moving convection cell flow ($c_1=0.521,c_2=3.7\times10^{-5}$) (b) fixed convection cell flow ($c_1=-0.256,c_2=6.2\times10^{-5}$) and (c) sinusoidal shear flow ($c_1=0.945,c_2=1.6\times10^{-3}$).}	
\end{figure}

This strong dependence of the steady state on the fragmentation process can also clearly be seen by how $\alpha_\infty$ changes with the coagulate strength  $\gamma$ (see Fig. \ref{fig:average_2}). We note that the shear forces in the convection flows and the sinusoidal shear flow have a different magnitude, as seen in Fig. \ref{fig:flow_gradient}. Therefore three different values of the coagulate strength parameter $\gamma$ need to be chosen to yield a size distribution that is not only determined by size-limiting fragmentation. For the convection flows the coagulate strength $\gamma$ needs to be approximately a factor $3$ larger than for the sinusoidal shear flow. 

It is clear that $\alpha_{\infty}$ increases with $\gamma$, because particles become more resistant to shear. The exact functional relationship is however not trivial. A first qualitative estimate of the shape of this $\alpha_{\infty}(\gamma)$ curve can be derived by assuming that over one period of the flow the particles experience an 'average shear'
\begin{equation}
\bar{G} = \frac{1}{T}\int\limits_0^T\hspace{-0.1cm}dt\int\limits_{D}\hspace{-0.15cm}d{\bf x}~p({\bf x},t)G({\bf x},t)~, 
\end{equation}
where $G({\bf x},t)$ is the modulus of the local velocity gradient, $p({\bf x},t)$ is the distribution of particles and $D$ is the unit square domain. From Eq. (\ref{eq:splitting_drop}) we then get for the average critical size at this velocity gradient
\begin{equation}\label{eq:alpha_crit}
 \bar{\alpha}_{crit}=\bar{G}^{-3}\gamma^3~.
\end{equation}
Particles that exceed this size will therefore typically break up during one period of the flow. Since particles break into two parts due to shear, the average size would then be $\alpha_{\infty}\geq\bar{\alpha}_{crit}/2$. The average shear $\bar{G}$ is, however, 
somewhat complicated to estimate. It would have to be calculated as a mean over the positions of all particles in the flow at a given time. Additionally, how much larger than the critical size particles get before they break up depends on the coagulation probabilities, and therefore also on the local concentrations of particles. The exact dependency of $\alpha_{\infty}(\gamma)$ is therefore not easily calculated. What can be seen from Eq. \eqref{eq:alpha_crit} is however that the average size is expected to scale with $\gamma$ as
\begin{equation}\label{eq:alpha_infty}
 \alpha_{\infty} \propto \gamma^3~.
\end{equation}

This dependence is expected to hold for all values of $\gamma$ and $\alpha_{\infty}$, where shear fragmentation dominates. A fit with Eq. \eqref{eq:alpha_infty} for the different flows is shown in Fig. \ref{fig:average_2} and for lower values of $\gamma$ the fits agree very well with the simulation results. It can be seen that for higher values of $\gamma$, when size-limiting fragmentation becomes important, the $\alpha_{\infty}(\gamma)$ curves deviate from this estimate and converge towards the limiting value $\alpha_{\infty}^{(lim)}$ (see Fig. \ref{fig:average_2}). This result demonstrates how the steady state depends very strongly on the fragmentation process. However, the different proportionality constants for Eq. \eqref{eq:alpha_infty} still depend on the flow and can also depend on the spatial distribution of the particles, since different regions of the flow might exhibit different shear. 

\begin{figure}[htb]
		\centering
		\includegraphics[width=0.46\textwidth]{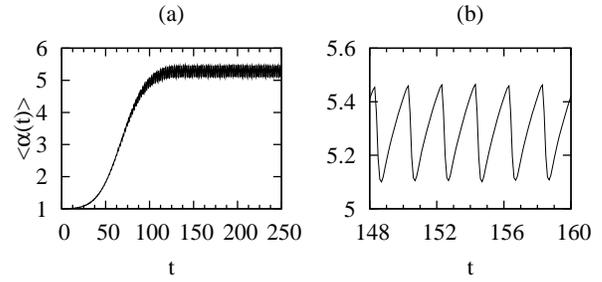}
 		\caption{\label{fig:average_1} In the case of shear fragmentation the steady state is not static, instead it fluctuates with the period of the flow: (a) average size class vs. time for the fixed convection flow and (b) close-up of (a).}	
\end{figure}

Finally, we mention some further results from our model. First, for all flows the steady state in the case of shear fragmentation is not static, instead due to the periodic time dependence of the flows the steady state also varies periodically over time. This is very clear for example for the fixed convection flows when looking at the average size class index $\left<\alpha(t)\right> = \sum_{i=1}^{30}\alpha_{i} N_{\alpha_{i}}(t)/N(t)$, where $N_{\alpha_i}$ denotes the number of particles in size class $\alpha_i$ (Fig. \ref{fig:average_1}).

Such a time dependence of the average particle size, and therefore of the whole particle size distribution, can have important physical consequences. For example the settling of coagulates in the ocean, which is an important part of the biological carbon pump in the ocean, is greatly affected by the size distribution of the coagulates. In coastal areas, where the fluid may be periodically forced by the tides, such a time dependence of the distribution can greatly affect the settling rates and therefore the deposition of coagulates \cite{Lunau2006}.

Second, since the range of possible parameter values for our model is very large, we also mention the robustness of our findings with respect to the different parameters:

\begin{figure}[htb]
		\centering
		\includegraphics[width=0.46\textwidth]{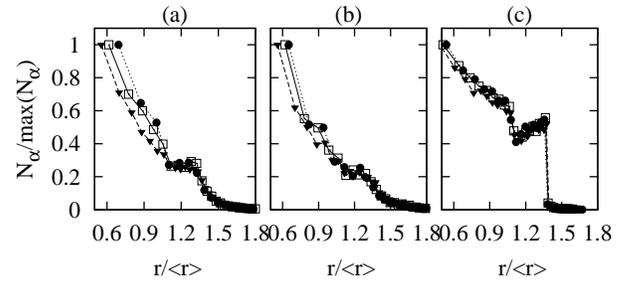}
 		\caption{\label{fig:hist_3} Normalized number of particles versus relative radius in steady state in the case of shear fragmentation for different values of the coagulate strength parameter $\gamma$. The normalized size distributions collapse onto a master curve that depends on the fragmentation mechanism and the shear forces in the flow for (a) moving convection cell flow ($\gamma=44$ (circle), $50$ (square), $56$ (triangle)), (b) fixed convection cell flow ($\gamma=42.5$ (circle), $45$ (square), $50$ (triangle)) and (c) sinusoidal shear flow ($\gamma=16$ (circle), $17$ (square), $18$ (triangle), $\beta=20/\pi$). }	
\end{figure}

(i) For each flow there is a certain range of the coagulate strength parameters
$\gamma$ where the size distribution for shear fragmentation is ``fully
developed''. By this we mean that $\gamma$ is large enough so that a
sufficiently large fraction of particles has left the smallest size class, but
$\gamma$ is small enough so that break-up due to size-limiting fragmentation
does not play a significant role. In this intermediate $\gamma$ range, where the particle size distribution is fully
developed, a scaling form
\begin{equation}
\frac{N_{\alpha}}{{\rm max}(N_{\alpha}) } = f \left( \frac{r}{\left<r\right>} \right)
\label{eq:scaling}
\end{equation}
 is found to hold, where $\left<r\right>$ represents the average radius. Note that the form of the size distribution is independent of $\gamma$. All distributions in this parameter range collapse then onto a single master curve. 

While this scaling form is independent of the parameters of the coagulation and of the coagulate strength, the difference between the different flows remains. More specifically, the scaling form $f$ changes when shear forces in the flow or the fragmentation mechanism, for example the distribution of fragments, is varied.

(ii) When investigating cases with different total mass $M$, we find that  for size-limiting fragmentation $\alpha_{\infty}$ is largely independent of $M$. For shear fragmentation with $M<3\times 10^{6} m_1$, $\alpha_{\infty}(M)$ increases approximately linearly with $M$, while for higher values a saturation of $\alpha_{\infty}(M)$ sets in, which is due to the fact that size-limiting fragmentation dominates in this case.

(iii) By considering other initial particle size distributions than mentioned above, for example any single size class with $\alpha>1$ or a uniform distribution of sizes, and keeping the total mass $M$ fixed, the asymptotic state is found to be independent of the chosen initial distribution for each flow and for both fragmentation rules. 

(iv) We also investigated the role of the number of size classes and found that in the chosen range of $\gamma$
values the size distributions for shear fragmentation are not influenced by the number of size classes.

(v) The effect of the number of new particles formed by fragmentation has been
considered. For instance, the distributions of particles for ternary
fragmentation are similar to the ones for binary splitting and only show a
slight shift towards smaller size classes \footnote{A change of the
fragmentation rule has, however, important consequences. By replacing the rule of uniform distribution for the possible size classes after fragmentation by the rule that coagulates always split into two halfs of similar sizes,
makes the size distribution exponential, practically over all size classes.}.

\section{Summary}
\label{sec:conclusions}
We discussed the formation of a steady state size distribution in a
coagulation fragmentation process of inertial particles transported by different flows. Our most important finding is that fragmentation rather than coagulation is the dominating process for the steady state size distribution. For size-limiting fragmentation we found almost no differences in the shape of the steady state size distribution for various flows. Even in flows with great differences in coagulation rates, e.g. due to differences in local particle concentrations, particle size distributions remained very similar. For the case of shear fragmentation differences in the shape of the size distribution for the different flows appeared. It was found that these were due to differences in the spatial and temporal variation of the shear, which in turn affected the fragmentation.

We have shown that an individual particle based modeling approach is able 
to reflect typical properties of coagulation and fragmentation processes of inertial particles. 
The appearance of a steady state is demonstrated. We outlined some of the differences in the convergence to the steady state and the particle size distribution that can result from different types of fragmentation and flow. Altogether, our results suggest that coagulation dominates different time spans of the process than fragmentation. While coagulation is most important for the transients in the beginning, the steady state size distribution is mainly determined by fragmentation. As a consequence the spatial distribution of particles plays only a transient role. The underlying flow is important for the steady state in the case of shear fragmentation, as the spatial and temporal variation of the shear can greatly influence the fragmentation rates.
 
The generalization to a fully three-dimensional system is straightforward. The relaxation towards the steady state would slow down, due to the decreased probability of collisions. However, our conclusions regarding the dependence of the steady state size distribution on the fragmentation remain valid.

\section{Acknowledgments}
The authors thank A. Alldredge, P. Franks, S. Malinowski, M. Maxey, J. Maerz, E. Villermaux and L.-P. Wang for useful discussions and suggestions. We acknowledge the support of the Hungarian Science Foundation under the contract OTKA T72037. 
\label{sec:acknowledgements}


\begin{thebibliography}{99}

\bibitem{Nishikawa2001}
T. Nishikawa, Z. Toroczkai and C. Grebogi, Phys. Rev. Lett. {\bf 87}, 038301 (2001).

\bibitem{Benczik2002} 
 I.J.~Benczik, Z.~Toroczkai, and T.~T\'el, Phys. Rev. Lett. {\bf 89}, 164501 (2002).

\bibitem{Lopez2002}
C. Lopez, Phys. Rev. E {\bf 66}, 027202 (2002).

\bibitem{Cartwright2002}
J.H.E. Cartwright,  M.O. Magnasco, O. Piro  and I. Tuval, Phys. Rev. Lett. {\bf 89}, 264501 (2002).

\bibitem{Do2004}
 Y.~Do and Y.C.~Lai, Phys. Rev. E {\bf 70}, 036203 (2004).

\bibitem{Benczik2006}
I.J. Benczik, G. Karolyi, I. Scheuring and T. T\'el, Chaos {\bf 16}, 043110 (2006).

\bibitem{Vilela2007}
R.D.~Vilela and A.E.~Motter, Phys. Rev. Lett. {\bf 99}, 264101 (2007).

\bibitem{Sapsis2008}
G.~Haller and T.~Sapsis, Physica D {\bf 237}, 573 (2008).


\bibitem{Shaw2003}
R.A. Shaw, Ann. Rev. Fluid Mech. {\bf 35}, 183 (2003).

\bibitem{Falkovich2007}
G. Falkovich and A. Pumir, J. Atmos. Sci. {\bf 64}, 4497 (2007).

\bibitem{Jaczewski2005}
A. Jaczewski and S.P. Malinowski, Q. J. R. Meteorol. Soc.
{\bf 131}, 2047 (2005).

\bibitem{Wilkinson2005}
M. Wilkinson and B. Mehlig, Europhys. Lett. {\bf 71}, 186 (2005).

\bibitem{Bec2005}
J. Bec, A. Celani, M. Cencini, and S. Musacchio, Phys. Fluids {\bf 17}, 073301 (2005).

\bibitem{Calzavarini}
E. Calzavarini, M. Cencini, D. Lohse,  and F. Toschi, Phys. Rev. Lett. {\bf 101}, 084504 (2008).

\bibitem{Maxey1986}
M.R. Maxey and S. Corrsin, J. Atmos. Sci. {\bf 43}, 1112 (1986).

\bibitem{Maxey1987a}
M.R. Maxey, J. Fluid Mech. {\bf 174}, 441 (1987).

\bibitem{Wilkinson2007}
M. Wilkinson, B. Mehlig, S. \"Ostlund and K.P. Duncan, Phys. Fluids {\bf 19}, 113303 (2007)

\bibitem{falko_nat02}
 G. Falkovich, A. Fouxon and M.G. Stepanov, Nature {\bf 419}, 151 (2002).

\bibitem{Zhou2001}
Y. Zhou, A.S. Wexler and L.-P. Wang, J. Fluid Mech. {\bf 433}, 77 (2001)

\bibitem{Wang2000}
L.-P. Wang, A.S. Wexler and Y. Zhou, J. Fluid Mech. {\bf 415}, 117 (2000)



\bibitem{Wilkinson2008}
M. Wilkinson, B. Mehlig and V. Uski, Astrophys. J. Supplement Series {\bf 176}, 484 (2008)

\bibitem{Zahnow2008_2}
J.C. Zahnow, R.D. Vilela, U. Feudel and T. T\'el, Phys. Rev. E {\bf 77}, 055301(R) (2008).

\bibitem{Medrano2008}
R.O. Medrano, A. Moura, T.T\'el, I.L. Caldas and C. Grebogi, Phys. Rev. E {\bf 78}, 056206 (2008).

\bibitem{Pruppacher1997}
H.R. Pruppacher and J.D. Klett, {\it Microphysics of Clouds and Precipitation}, (Kluwer Academic Publishers, Dordrecht, 1997).

\bibitem{Alldredge1990} 
A.L. Alldredge, T. Granata, C. Gotschalk, and T. Dickey, Limnology and Oceanography {\bf 35}, 1415 (1990).

\bibitem{Viller2008}
E. Villermaux, Annu. Rev. Fluid. Mech. {\bf 39}, 419 (2007) 

\bibitem{Thomas1999}
D.N. Thomas, S. Judd and N. Fawcett, Wat. Res. {\bf 33}, No. 7, 1579 (1999).

\bibitem{Smoluchowski1917}
M. Smoluchowski, Zeitschrift f\"ur physikalische Chemie, {\bf 92}, 129 (1917). 

\bibitem{Bec2003}
J. Bec, Phys. Fluids {\bf 15}, L81 (2003);
Phys. Fluids {\bf 17}, 073301 (2005);

\bibitem{Kundu} 
P.K. Kundu, I.M. Cohen, {\em Fluid Mechanics}, Fourth Edition, Academic Press, 2008




\bibitem{Maxey1983}
M.R. Maxey and J.J. Riley, Phys. Fluids {\bf 26}, 883 (1983)

\bibitem{Auton1988}
T.R. Auton, J. Hunt and M. Prud'homme, J. Fluid. Mech. {\bf 197}, 241 (1988)

\bibitem{Michaelides1997}
E.E. Michaelides, J. Fluids Eng. {\bf 119}, 233 (1997)

%

\bibitem{Happel1983}
J. Happel and H. Brenner, {\it Low Reynolds number hydrodynamics}, (Martinus Nijhoff Publishers, The Hague, 1983)

\bibitem{JTWG1997}
I.M. J\'anosi, J.O. Kessler and V.K. Horv\'ath,
Phys. Rev. E {\bf 58}, 4793 (1998)

\bibitem{concentration}
For the cases studied here, typical distances between particles are $\geq50a_1$.



\bibitem{Taylor1934}
G.I. Taylor, Proc. Roy. Soc. A {\bf 146}, 501, (1934)

\bibitem{Delichatsios1975} 
M. Delichatsios, Phys. Fluids {\bf 18}, 622, (1975)

\bibitem{Chandrasekhar}
S. Chandrasekhar, {\it Hydrodynamic and Hydromagnetic Stability}, (Oxford University Press, 1961)

\bibitem{Zahnow2008}
J.C. Zahnow and U. Feudel, Phys. Rev. E {\bf 77}, 026215 (2008)

\bibitem{Maxey1987}
M.R. Maxey, Phys. Fluids {\bf 30}, 1915, (1987)

%
%
%
%

\bibitem{Nishikawa2002}
T. Nishikawa, Z. Toroczkai, C. Grebogi and T. T\'el, Phys. Rev. E {\bf 65}, 026216 (2002)

\bibitem{Liu1994}
M. Liu, F. Muzzio and R. Peskin,
Chaos Sol. Fract {\bf 4}, 869 (1994)

\bibitem{Pierre1994}
R.T. Pierrehumbert,
Chaos Sol. Fract {\bf 4}, 1091 (1994)


\bibitem{Vilela2007_2}
R.D. Vilela,  T. T\'el, A.P.S. de Moura and C. Grebogi, Phys. Rev. E {\bf 75}, 065203(R) (2007) 

\bibitem{Hockney1981}
R.W. Hockney and J.W. Eastwood, {\it Computer Simulations using Particles}, (McGraw-Hill, New York, 1981)

\bibitem{Lunau2006}
M.~Lunau, A.~Lemke, O.~Dellwig, M.~Simon, Limnology and Oceanography {\bf 51}, 847 (2006)

\end{thebibliography}
\end{document}